\documentclass{IEEEtran}
\usepackage{upgreek}
\usepackage{graphicx}
\usepackage{subfigure}
\usepackage{amsmath,bm}
\usepackage{enumerate}
\usepackage[mathscr]{euscript}
\usepackage[square, comma, sort&compress, numbers]{natbib}
\usepackage{xcolor}
\usepackage{url}

\ifCLASSINFOpdf
\else
\fi

\begin{document}

\title{Mobile Power Network for Internet of Things without Battery Life Anxiety}

\long\def\symbolfootnote[#1]#2{\begingroup%
\def\thefootnote{\fnsymbol{footnote}}\footnote[#1]{#2}\endgroup}
\renewcommand{\thefootnote}{\fnsymbol{footnote}}
\author{Mingqing~Liu,
Qingwen~Liu,
Mingliang~Xiong,
Hao~Deng,
and Xinhe Wang

\thanks{
This work has been submitted to the IEEE for possible publication. Copyright may be transferred without notice, after which this version may no longer be accessible.}
}


\maketitle

\begin{abstract}
Similar to the evolution from the wired Internet to mobile Internet (MI), the growing demand for power delivery anywhere and anytime appeals for power grid transformation from wired to mobile domain. We propose here the next generation of power delivery network -- mobile power network (MPN) for wireless power transfer within a mobile range from several meters to tens of meters. At first, we overview the motivation for proposing MPN and present the MPN's concept evolution. Then, we report the MPN's supporting technologies, and particularly introduce resonant beam charging (RBC). As a long-range wireless power transfer (WPT) method, RBC can safely deliver multi-Watt power to multiple devices concurrently. Meanwhile, the recent progress in RBC research has been summarized. Next, we specify the MPN's architecture to provide the wide-area WPT coverage. Finally, we present the MPN's application scenarios and discuss key issues in MPN. MPN can enable the ultimate mobility by cutting the final cord of mobile devices, realizing the ``last-mile'' mobile power delivery.
\end{abstract}


\IEEEpeerreviewmaketitle

\section{Introduction}
\label{sec:Introduction}
More recently, the Internet of Things (IoT), the interconnection of computing devices embedded in everyday objects, has gone from a futuristic concept to an actual industry concern. It is estimated that the number of IoT devices connected to the Internet is growing to billions. However, continuity is one of the top challenges of IoT as ensuring and extending battery life is an essential consideration for IoT devices. To deal with the above challenge, one solution is to optimize the power consumption of IoT devices, while the other is to power IoT devices wirelessly. As ultra-low power consumption technology is not 
feasible for devices such as smartphones, virtual reality  (VR)~/~augmented reality (AR) devices, unmanned aerial vehicles (UAV), etc., which are necessary parts of IoT, wireless power transfer (WPT) seems to be a possible way for leveraging permanent battery life of IoT devices. 

\begin{figure}[hbt]
	\centering
	\includegraphics[width=3in]{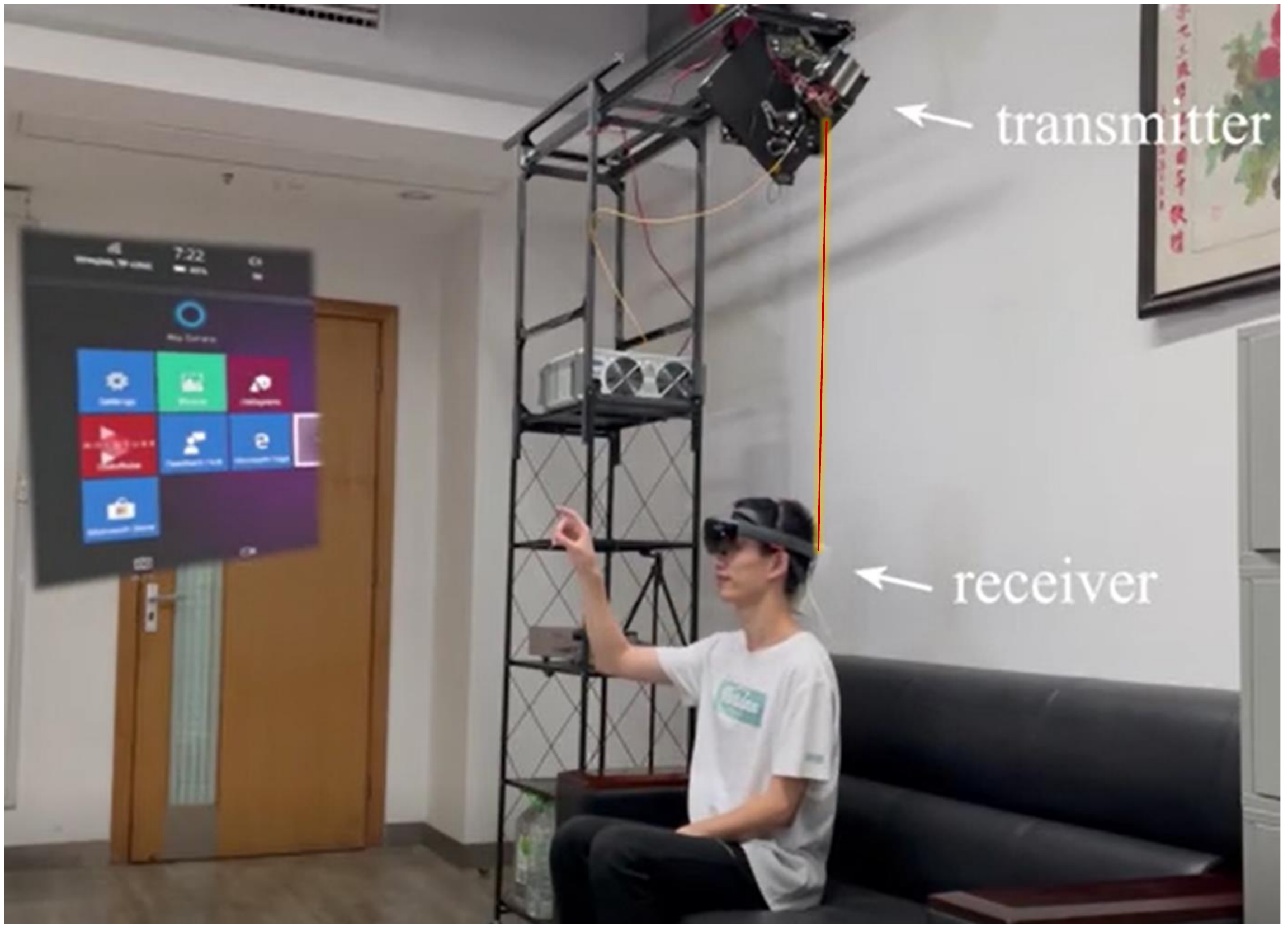}
	\caption{RBC Application in Wireless Charging AR Glasses Over the Air}
	\label{fig:AR}
\end{figure}

Since the concept of IoT was proposed, new developments have been shaking up old and new industries, where the smart factory is one of the critical demonstrations. Especially under the influence of COVID-19, the deployment of IoT not only realizes remote monitoring solutions to help improve manufacturing efficiency and workshop automation to assist personnel but also helps improve employees' health monitoring levels. In terms of transportation and logistics, intelligent logistics tracking relying on IoT technologies dramatically improves operational efficiency and provides a foundation for the pharmaceutical supply chain. Besides, smart homes, smart offices, intelligent medical, etc., have brought enormous demands and markets. Moreover, smartphones, VR/AR devices, and wearable devices that are closely related to everyone every day consist of the IoT as a considerable part. However, cutting the last electric wire 
while providing a permanent energy supply without batteries for IoT devices is a vision for future IoT. To realize the actual connectivity of everything, WPT is a promising supporting method.

The existing WPT technologies include inductive coupling, magnetic resonance, capacitive coupling, etc., which are capable of providing watt-level power over centimeter-level distance, and radio frequency (RF) charging, laser charging, etc., which can transfer wireless power over several meters distance with milliwatt-level power limited by safety requirement or with beam-steering control for alignment. After more than ten years of development in academia and industry, WPT is still facing a bottleneck in simultaneously achieving high transmission efficiency and self-alignment between the WPT transmitter and the receiver. Resonant beam charging (RBC), also known as distributed laser charging (DLC), can simultaneously realize high-power, long-range, and mobile power transfer for IoT devices without the need for 
mechanical alignment control as in Fig.~\ref{fig:AR}, which is capable of solving the ``last-mile'' mobile power delivery problem and providing power anytime and anywhere. Various WPT can be applied in various scenarios which enable mobile power transfer in the energy domain similar to wireless information transfer (WIT) in the information domain~\cite{Li2014Cellular,6G}. However, to realize ``mobile power delivery'' anywhere and anytime, providing wide-area mobile charging services is vital. On the one hand, the power grid is accessed to supply energy in the mobile realm. On the other hand, the Internet is connected to provide wide-area services such as access, resource scheduling, deployment, etc. 

	\begin{figure}[t]
	\centering
	\includegraphics[width=3.5in]{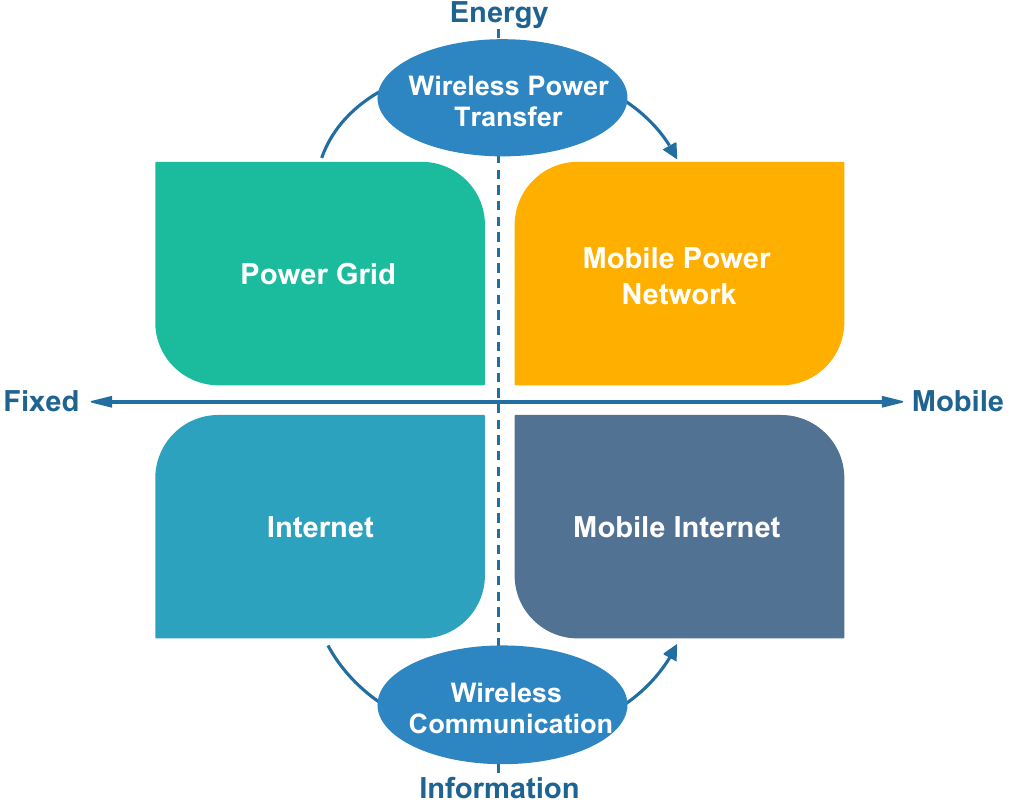}
	\caption{Concept Evolution}
	\label{fig:Conceptual-Positioning}
\end{figure}

 Given such motivations, we propose the paradigm of a mobile power network (MPN), to extend power delivery to the mobile realm. As shown in Fig.~\ref{fig:Conceptual-Positioning}, the concept of MPN is derived from the evolvement of the Internet to mobile Internet (MI). Wireless communications enable the evolution from the Internet to the MI. MI realizes the information dissemination over the air and solves the ``last-mile'' information delivery problem. Mobile information delivery anywhere and anytime brings a free lifestyle of study, work, entertainment, and business. Similarly, MPN is presented to transfer power over the air and solve the ``last-mile" power delivery problem based on WPT technologies. MPN can provide wireless power to mobile terminals which is similar to wireless information delivery in MI. Therefore, MPN is the extension of the power grid for mobile power delivery. People will get mobile power no longer restricted by safety, time, space, and other limitations, bringing anywhere and anytime power supply into reality~\cite{6G}.

\begin{figure}
	\centering
	\includegraphics[width=3.5in]{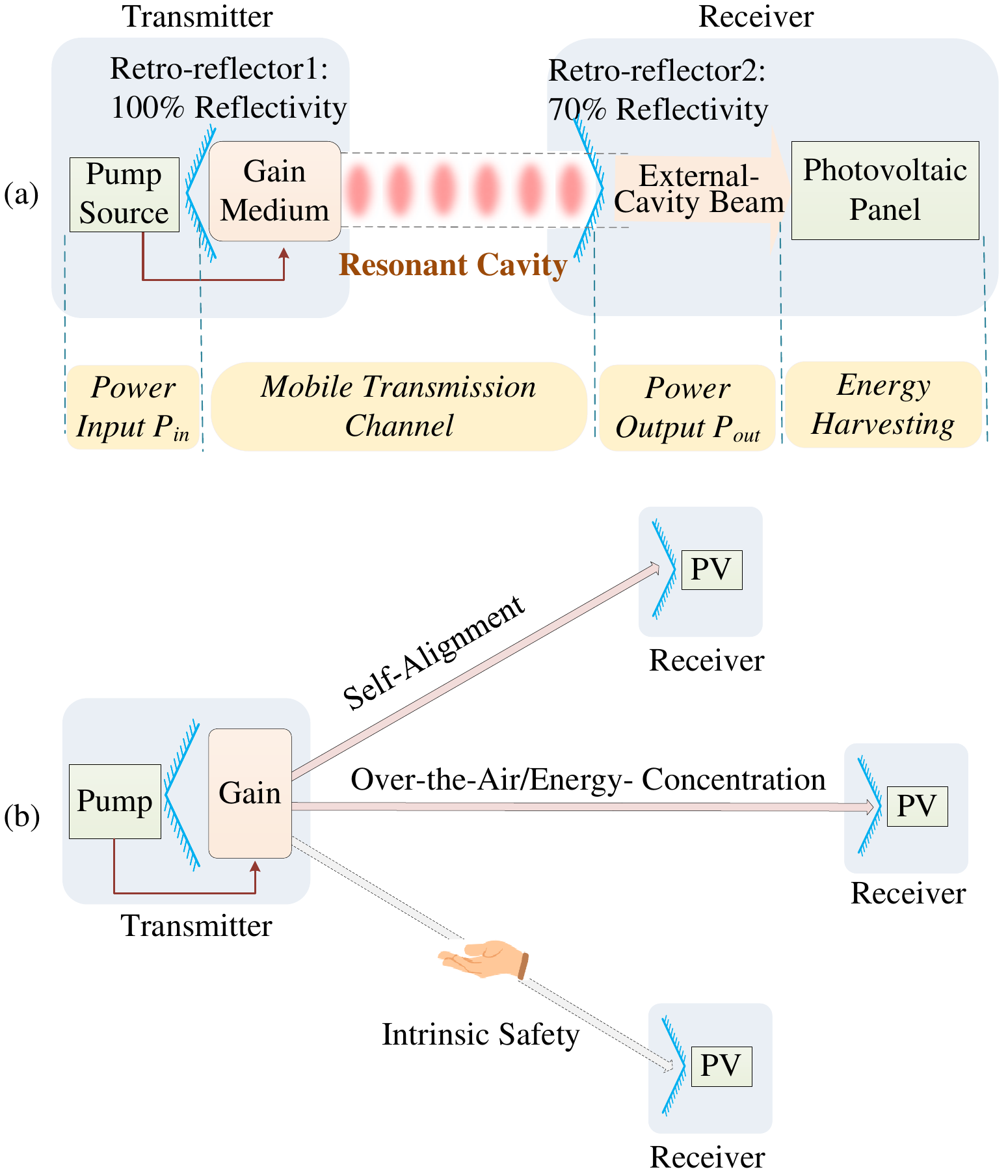}
	\caption{Resonant Beam Charging Principle and Features}
	\label{fig:RBC-Features}
\end{figure}

\section{Supporting Technologies}
\label{sec:SupportingTechnologies}
WPT technologies are the foremost supporting technologies to realize MPN. Nikola Tesla invented the ``Tesla coil'' for the initial wireless power transmission test one hundred years ago. In recent years, the development of wireless charging technologies has made great strides~\cite{lu2016wireless}. WPT technologies that draw significant attention in industry and academia include inductive coupling, magnetic resonance, capacitive coupling, RF charging, laser charging, etc. However, the above WPT technologies face the challenges of simultaneously satisfying mobility, high power, and safety requirements in MPN applications. Thus, we will particularly introduce the RBC technology, which can safely transfer watt-level power over several meters~\cite{Liu2016Charging}.

\subsection{Existing WPT Technologies}
WPT mainly uses an electromagnetic field as a carrier to transmit energy in open space, including near-field and far-field technologies. Near-field technologies, using the non-radiative magnetic field (also known as an evanescent wave) of the transceiver coil to realize the coupling and transfer of energy between coils, are mainly divided into two coupling modes: magnetic induction and magnetic resonance. The well-known wireless charging pads generally adopt the above near-field technologies. Far-field technologies, which use radiated electromagnetic fields (also called electromagnetic waves) to transmit energy from the transmitter to the receiver, are broadly divided into the following two kinds: non-directional and directional radiation. Among them, the transmission carriers of non-directional far-field WPT are non-directional RF and solar energy, while the transmission carriers of directional far-field WPT are directional RF and laser. 

Near-field WPT is generally safe for humans while the charging distance and tight alignment are the defects restricting mobility. To this end, near-field WPT can hardly empower applications such as metaverse in the next generation of networks. Recently, efforts in both academia and industry are turning to focus on far-field WPT, which can achieve tens of meters transmission distance. Meanwhile, research on safety and high-power transfer has been explored in far-field WPT. Vikram Iyer \textit{et al.} used protective beam technology to transmit energy through traditional lasers and provide radiation safety guarantees~\cite{2018Charging}. Xu Zhang \textit{et al.} published a secure energy harvesting technology in the Wi-Fi frequency band, using MoS2 enhanced silicon rectifier diode antennas to achieve $0.156$mW power wireless charging at a distance of $1$m~\cite{2019Two}. Additionally, the theoretical research and experiment of RBC have verified the wireless charging with a power of $2$W at a distance of $2.6$m~\cite{Wangw}, demonstrating its potential in providing wireless power supply services in the IoT era.

\begin{figure}[t]
	\centering
	\includegraphics[width=3.3in]{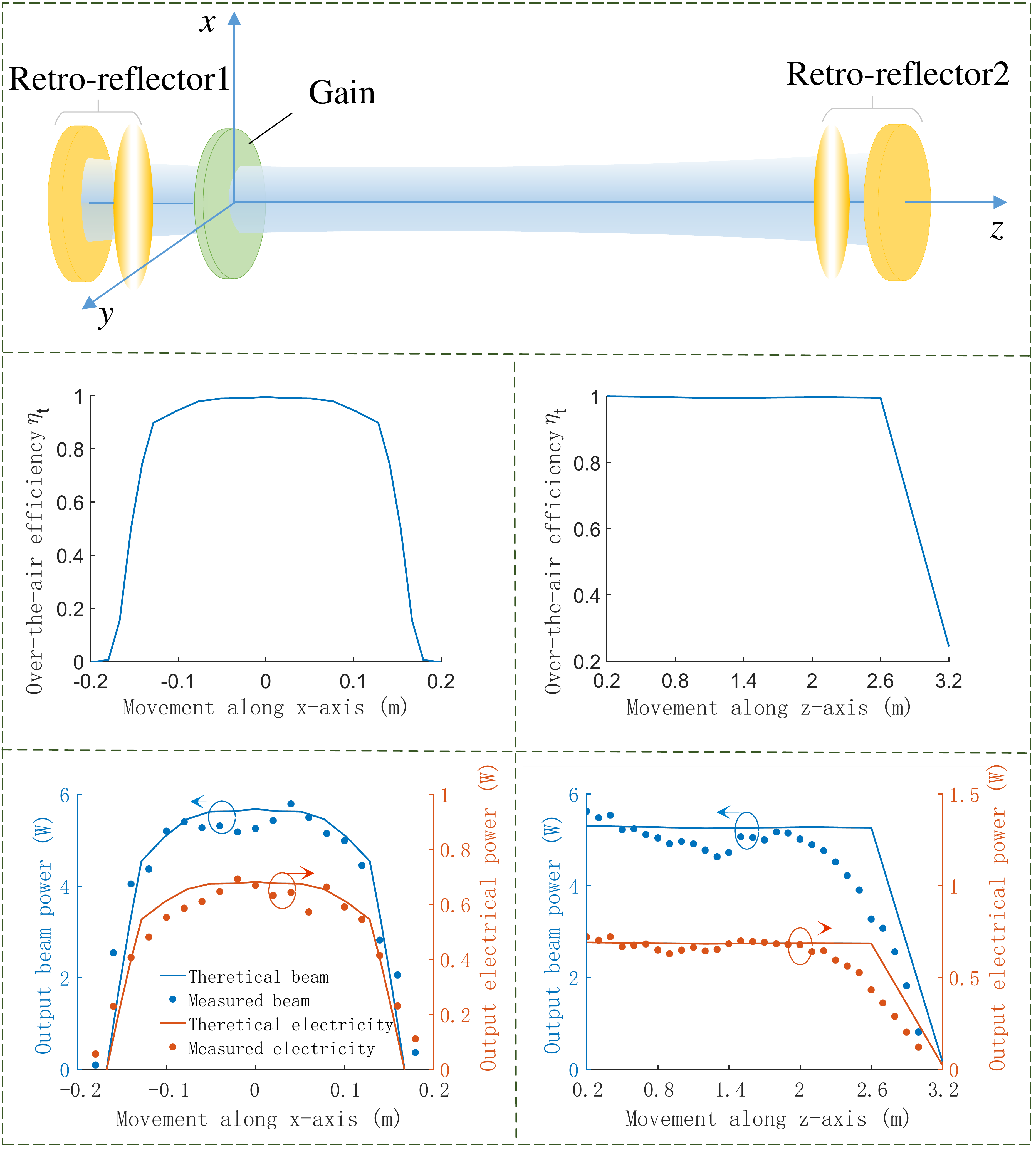}
	\caption{Simulation and Experimental Results (Figs. 7-10 in Ref.~\cite{testbed})}
	\label{fig:testbed}
\end{figure}

\subsection{Resonant Beam Charging}

\subsubsection{RBC Principle and Features}
Figure~\ref{fig:RBC-Features} (a) illustrates the RBC system diagram. A retro-reflector, a gain medium, and the pump source are integrated at the transmitter, while the other retro-reflector and a photovoltaic (PV) panel are integrated at the receiver. The transmitter and the receiver jointly form a resonant cavity to generate a resonant beam, through which the power is transferred wirelessly. 

The retro-reflector ensures that the resonant beam entering from a wide range of angles can be returned back to its incoming direction~\cite{Liu2016Charging}. In this system, the pump source at the transmitter first stimulates the gain medium to generate a resonant beam. The beam reflected by the two retro-reflectors is folded back into the gain medium multiple times to be amplified. Thus, a continuous resonant beam of stimulated radiation exists within the resonator (i.e., the transmitter and receiver). Subsequently, the resonant beam partially passes through the partially reflective retro-reflector 2 at the receiver to form a laser beam. Finally, the laser beam is converted into the output electrical power with a photovoltaic (PV) panel, which can be readily used to charge electronic devices.

The unique features of RBC's mobile transmission channel rely on the generation principles of lasers: given a certain pump source to the gain medium and a resonant cavity within which beams can propagate back and forth, resonant beams in the mobile transmission channel can be formed. Thus, in the RBC system, the alignment between the transmitter and the receiver is automatically generated so that the receivers can still be charged while moving~\cite{mobility}. Moreover, any foreign object that enters the line of sight, i.e., the beam path, will block the photons in the path, disrupt positive feedback for resonance, and automatically shut off the resonant beam. Thus, the requirements of high-power transmission and safe charging can be guaranteed. In addition, the resonant beam is essentially an intra-cavity laser inheriting the advantages of lasers such as energy-concentrated narrow-beam transmission. Thus, the transmission efficiency of the resonant beam is significantly high compared to RF supporting long transmission distances. Moreover, RBC also has the characteristics of high-power, concurrent-charging, and compact size as specified in~\cite{Liu2016Charging}. The receiver can be integrated into various IoT devices, which leads to the feasibility of practical implementation.

\subsubsection{Recent Progress in RBC Research}
Since RBC has been proposed~\cite{Liu2016Charging}, RBC has made great strides in both theoretical research and testbed establishment. We summarize the RBC progress in the following three aspects with the consideration of IoT architecture as physical, access, and networking.

In terms of the physical layer, we investigate the principle and features of the RBC deeply. At first, core features of RBC including long transfer distance, mobility, and inherent safety have been studied~\cite{mobility,safety}. RBC's theoretical transmission distance can reach hundreds of meters. The mobility mechanism of the RBC has been revealed, with which the simultaneous lightwave information and power system (SLIPT) extended by RBC is demonstrated to have the capability of supplying over $3$W power over $2$m distance within over $20^{\circ}$ field of view (FoV)~\cite{mobility}. Moreover, the inherent safety has been presented and verified through theoretical analysis that $1$W power can be transferred over $5$m distance under the skin-safe regulations for laser products~\cite{safety}. Besides, we have built a testbed that verified the watt-level power transfer over meter-level distances within a large FoV with the premise of human safety~\cite{testbed} as in Fig.~\ref{fig:testbed}. The smartphone can be charged over-the-air while it is moving. Over-the-air transmission efficiency of RBC is nearly 100\%, and the mobile transmission channel remains with the receiver's movement within a specific coverage.

In terms of the access layer, to avoid power waste and accidental dangers, the adaptive resonant beam charging (ARBC) design is presented for adaptively controlling the transmitting power from the transmitter according to the charging requirements of the receiver. Consequently, the efficiency of RBC has also been enhanced by significantly cutting energy waste. Based on RBC, for better charging services, a wireless power scheduling algorithm according to battery charging profile for fairly keeping all devices working as long as possible is proposed. Moreover, the scheduling algorithm for earning maximization with quality of charging service guarantee, and the TDMA-based WPT scheduling algorithm for efficient WPT in ARBC are put forth~\cite{2fang2018}. These efforts are to drive the progress and implementation of wireless charging services.
\begin{figure*}
	\centering
	\includegraphics[width=5in]{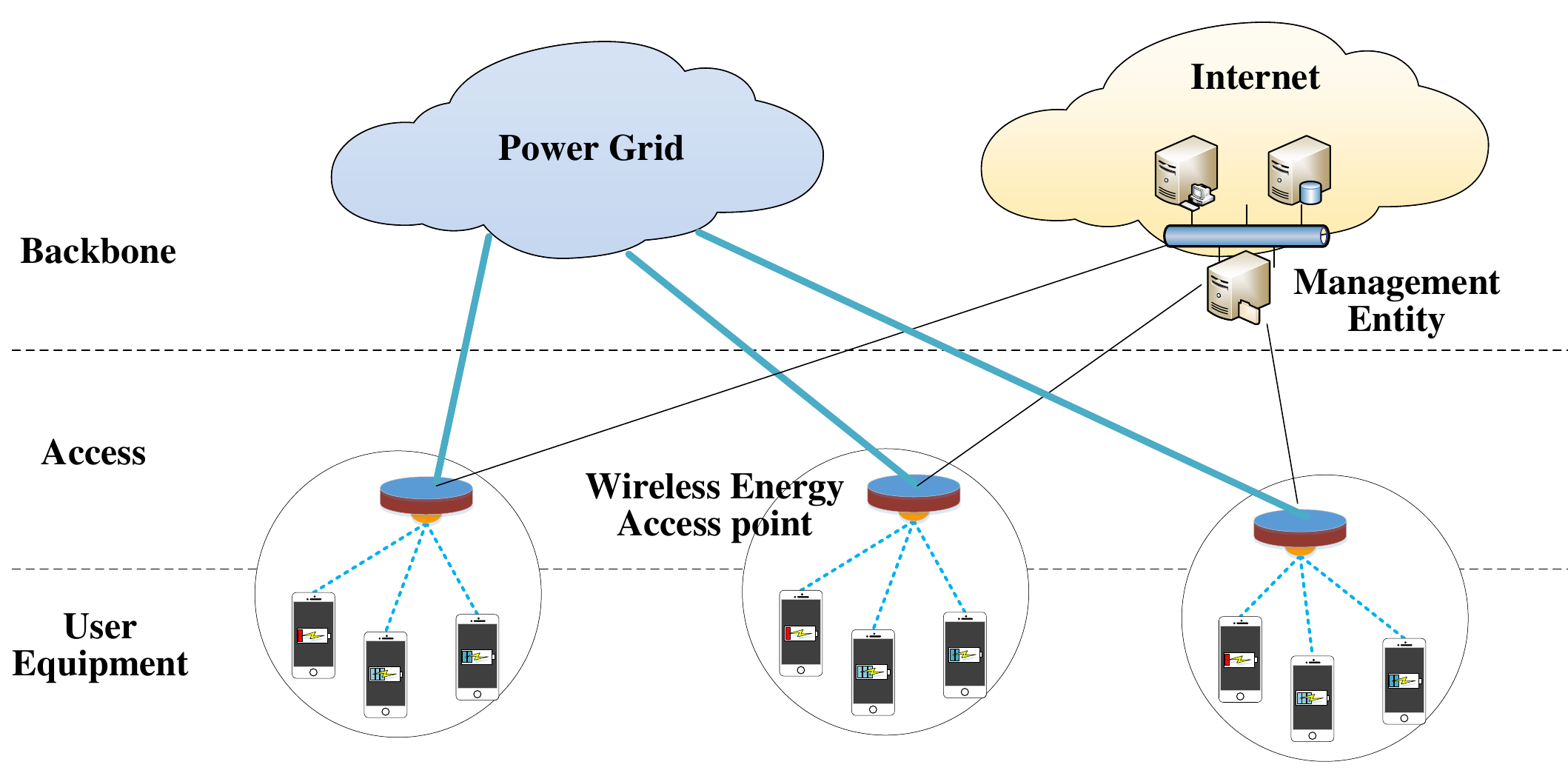}
	\caption{MPN Architecture}
	\label{fig:Star-Topology}
\end{figure*}

In terms of the networking layer, a wide-area deployment algorithm is proposed for providing wide-area wireless charging services, which also provides the foundation of a large-scale power schedule. Besides, a wireless charging application has been developed to control and visualize users' charging time, charging fees, and charging performance.

Above all, research on RBC basically covers all aspects of network construction, contributing to the design and implementation of MPN. Moreover, many researchers have joined us to carry out RBC-related theoretical and experimental research, further confirming the feasibility of RBC~\cite{Hangguo,Sheng21,liu2022large}.

\section{Mobile Power Network Architecture}
\label{sec:MobileEnergyInternetArchitecture}

The MPN architecture requires not only far-field wireless charging technologies but also near-field wireless charging technologies for their respective application scenarios. For example, inductive coupling and magnetic resonance technology can be applied in the MPN infrastructural network to supply power for home appliances and electric vehicles. Laser charging or RF charging may be used in the MPN-powered WSNs. Based on comparing these WPT technologies, we can find that RBC is the MPN's enabling technology for mobile applications~\cite{Liu2016Charging}. We designed the mobile power network (MPN) architecture based on RBC to provide wide-area mobile charging services. Relying on the features of RBC technology, one transmitter can charge several receivers simultaneously, while one receiver can be served by several transmitters. Thus, the control functions such as access control, scheduling control, and mobility management should be provided by the MPN.

\subsection{Major Components}
In Fig.~\ref{fig:Star-Topology}, the MPN architecture is separated into two layers: the access layer and the backbone layer. The access layer takes charge of wireless power transfer for the user equipment. Moreover, the entities in the access layer can operate control of user access, scheduling, and channel switching. In the backbone layer, wireless power access points are connected to the power grid to get source power. Furthermore, Internet access is significant in providing signaling transmission for centralized  management. The major components in the MPN architecture are as follows:

\setcounter{subsubsection}{0}\subsubsection{User Equipment} User equipment (UE) represents devices such as smartphones that acquire mobile charging services by accessing MPN. UE has a built-in RBC receiver to receive the wireless power and then convert the wireless power to electrical power so that UE can be charged wirelessly. In addition, UE should be able to exchange information with the wireless power/data access point to support the control functions they needed.

\subsubsection{Wireless Power/Data Access Point} Wireless power/data access point (hereinafter called access point, AP) provides wireless power for receivers by a built-in RBC transmitter. It connects to the fixed power grid for the source power and accesses to the Internet for centralized management. AP assists UE in accessing and ensuring mobile charging quality. It also assists resource scheduling of MPN. According to the signaling from the management entity, AP relies on the power controller to determine the power allocation in a multi-user scenario. Moreover, AP is responsible for switching power transfer channels when UE moves from one AP charging area to another like handover in mobile communication networks.

\subsubsection{Management Entity} It is responsible for centralized information processing such as transaction data, real-time status, and resource management in the network, to provide the best decision for power transmission, access control, and resource allocation. It contains the functions of authentication, authorization, accounting (AAA), and location-based service (LBS) for mobility management. Furthermore, it takes charge of handover control for wide-area coverage. It also provides the application interfaces for value-added enhancement.

\begin{figure*}
	\centering\includegraphics[width=6.6in]{Fig3.pdf}
	\caption{MPN Application Scenarios}
	\label{fig:Scenarios}
\end{figure*}
\subsection{Features and Opportunities}
\label{sec:Features}
The wireless charging networks relying on near-field WPT technologies can only be deployed in a small area with limited mobility and safety~\cite{chargernetwork}. The concept of MPN, which is based on RBC as enabling technology, can highlight the characteristics of mobility and safety. In addition to the advantages of the RBC system mentioned earlier, the main features of MPN can be summarized as follows:

\setcounter{subsubsection}{0}\subsubsection{Mobility} RBC architecture ensures the receivers are charged while moving without the assistance of specific aiming or tracking~\cite{Liu2016Charging}. On the other hand, MPN enables WPT transmitters to be connected and centrally managed to provide wireless power in a wide range. MPN expands the power transfer over distance so that users can have a mobile WiFi-like experience to obtain wireless charging services with mobility.

\subsubsection{Safety} When encountering an obstruction within the resonant beam path in the RBC system, the resonation ceases at the speed of light since the photons are blocked by the obstacle. Thus, the power transfer is curtailed immediately without any software or decision-making circuitry involvement, which can ensure the coexistence of high power and safety. With the unique features of the RBC system, MPN can provide a wide-area wireless power supply with safety.

\subsubsection{Power} As the mobility-enabling technology, RBC can provide watt-level power over meter-level distance with the premise of safety. The power transferred by RBC is sufficient for charging mobile devices like smartphones. Thus, MPN is a high-power network that can support various high-power mobile charging scenarios.


In summary, MPN will bring convenience with enhanced charging experiences to human life in various aspects, such as medical treatment and smart homes. It breaks the development bottlenecks in dealing with the power supply problems for IoT devices. The simultaneous wireless information and power transfer (SWIPT) can be promoted with MPN. Furthermore, MPN will motivate new ideas to find solutions and possibilities for a more imaginative world. New business models will be built, and new lifestyles will be created.

\section{Application Scenarios and Discussions}
\label{sec:ApplicationScenarios}

The intelligent interconnection of cloud-edge-terminal infrastructure has become a significant development trend, such as smart-city IoT, industrial IoT, smart connected cars, etc. However, the intelligent and networked development of cloud-side-end infrastructure is facing various resource constraints such as bandwidth, computing power, storage, and energy consumption, which restricts the development of mobile-side intelligent interconnection. Nowadays, 
the landing of 5G technology and the research of 6G technology alleviate the bandwidth problem of mobile networking of smart terminals. Moreover, the development of cloud computing, edge computing, and collaborative computing technology alleviated the problem of computing power, while distributed storage technology (including blockchain technology) alleviated storage problems. However, in terms of energy consumption, there is still a lack of mature mobile power supply solutions for the mobile side. The proposed MPN is one of the technical systems that can deal with the above problems. We depict the MPN's application scenarios in this context as in Fig.~\ref{fig:Scenarios} and discuss critical issues to be solved.

\subsection{Applications}
A dividing line separates the fixed power grid and the Internet from MPN, while the right part shows the mobile essence of MPN: solving the mobile power transmission problems of the last few meters or tens of meters. Different from the existing fixed power grid, MPN focuses on wireless power transmission over the air and delivers power primarily depending on a variety of WPT technologies, so that power can be obtained without restrictions of fixed infrastructures. MPN includes the centralized control and management of WPT relying on mobile communication technologies. The power in MPN is initially provided by the fixed power grid, while the information such as the control signaling in MPN is upward transferred through the Internet. In the mobile realm, the application scenarios of MPN are as follows.
\setcounter{subsubsection}{0}\subsubsection{Smart Home}
Smart home devices can be supplied with wireless power through an MPN infrastructural network, where the wireless power transmitters can be embedded in light bulbs on the ceiling. The transmitter is connected to the Internet, and the client-server management system may be offered by the MPN supplier so that wireless power can be managed centrally and conveniently providing an integrated service experience. The network can also be deployed in public places such as coffee shops, airport terminals, and theatres to provide charging services covering the room. 

\subsubsection{Logistics} In industrial IoT scenarios where power lines are inconvenient to connect, logistics robots, power monitoring cameras, and other equipment can be powered by MPN. On the one hand, the non-interference RBC system can satisfy the requirements of safely supplying high power over long distances. On the other hand, networking allows edge intelligence as path planning, resource scheduling, deployment scheme, etc, can be provided through MPN.

\subsubsection{Extreme Environment} Wireless sensor network (WSN) is widely deployed with the demands of monitoring the environment in forests, deserts, or oceans. MPN can be applied to deal with the power endurance challenges of WSN, to save the cost of replacing a large number of batteries and avoid risks to humans. The powered WSN can self-determined the power resource schedule according to the architecture of MPN.

\subsubsection{Gatherings} The MPN ad-hoc network aims at providing wireless charging services in temporarily established outdoor situations. It can serve scenarios such as large gatherings, shared traffic, and emergency rescues. UAVs, integrated with the wireless power transmitter, can provide power to each other to ensure the stability and long-term functionality of the power supply network. Meanwhile, UAVs can serve devices with mobile charging requirements within their charging coverage. The remote management system controls the UAV-based power network.

\subsubsection{UAV Communication Networks} The UAV base station is expected to become a flexible and reliable communication base station (including for 5G), especially as an emergency plan under typhoons, landslides, earthquakes, natural disasters, and extreme conditions. MPN can be applied to assist UAV communication networks in dealing with power supply problems, where both the RBC transmitter and receiver can be embedded in the UAV. The specific base station provides power for them.

\subsection{Discussions}
From the application of MPN, we can summarize the key issues that MPN should focus on.
\setcounter{subsubsection}{0}\subsubsection{WPT with Safety, Mobility, and Long Distance} As illustrated above, MPN will be deployed in the human environment. Thus, ensuring safety is the primary condition that the WPT technology adopted in the MPN needs to meet. On the other side, to realize mobile power transfer in various scenarios, long-distance power transfer with the self-alignment of the WPT is required.

\subsubsection{Smart Grid Access} MPN is supposed to access the smart grid for obtaining convenient control as deployment, schedule, etc. However, compatibility issues in the process of smart grid access need to be resolved and feasible middleware needs to be designed and implemented. Besides, some access hardware designs may also be needed.

\subsubsection{Intelligent Deployment and Resource Scheduling} Random access of massive mobile terminals leads to dynamic changes in the temporal and spatial distribution of electricity demand, which poses challenges to MPN energy efficiency optimization and load balancing. Thus, we can adopt the behavioral cognition method of massive power consumption terminals and the prediction theory of power consumption terminal behavior based on information entropy, to investigate data-driven power supply optimization strategies.

\section{Conclusions}
\label{sec:Conclusion}
Driven by the demand of solving the power-hungry problem for mobile and IoT devices, we propose the mobile power network (MPN) based on wireless power transfer (WPT) technologies. We first present the MPN's concept evolution and application scenarios. We then introduce the MPN's supporting technology -- resonant beam charging (RBC), and demonstrate its features, performance,  and recent progress. Next, we present the MPN architecture to provide wide-area mobile charging services. Finally, we illustrate the application scenarios of MPN and discuss the challenges and open issues.
MPN will solve the ``last-mile'' mobile power delivery problem in the next generation of the Internet so that the ultimate mobility without ``battery life anxiety'' can be realized.

\bibliographystyle{IEEEtran}
\bibliography{Reference}

\begin{thebibliography}{10}
\providecommand{\url}[1]{#1}
\csname url@samestyle\endcsname
\providecommand{\newblock}{\relax}
\providecommand{\bibinfo}[2]{#2}
\providecommand{\BIBentrySTDinterwordspacing}{\spaceskip=0pt\relax}
\providecommand{\BIBentryALTinterwordstretchfactor}{4}
\providecommand{\BIBentryALTinterwordspacing}{\spaceskip=\fontdimen2\font plus
\BIBentryALTinterwordstretchfactor\fontdimen3\font minus
  \fontdimen4\font\relax}
\providecommand{\BIBforeignlanguage}[2]{{%
\expandafter\ifx\csname l@#1\endcsname\relax
\typeout{** WARNING: IEEEtran.bst: No hyphenation pattern has been}%
\typeout{** loaded for the language `#1'. Using the pattern for}%
\typeout{** the default language instead.}%
\else
\language=\csname l@#1\endcsname
\fi
#2}}
\providecommand{\BIBdecl}{\relax}
\BIBdecl

\bibitem{Li2014Cellular}
C.~Wang, F.~Haider, X.~Gao, X.~You, Y.~Yang, D.~Yuan, H.~M. Aggoune, H.~Haas,
  S.~Fletcher, and E.~Hepsaydir, ``Cellular architecture and key technologies
  for 5{G} mobile communication networks,'' \emph{IEEE Commun. Mag.}, vol.~52,
  no.~2, pp. 122--130, 2014.

\bibitem{6G}
K.~David and H.~Berndt, ``6{G} vision and requirements: Is there any need for
  beyond 5{G}?'' \emph{IEEE Veh. Technol. Mag.}, vol.~13, no.~3, pp. 72--80,
  Sept. 2018.

\bibitem{lu2016wireless}
X.~Lu, P.~Wang, D.~Niyato, D.~I. Kim, and Z.~Han, ``Wireless charging
  technologies: {F}undamentals, standards, and network applications,''
  \emph{IEEE Commun. Surveys Tuts.}, vol.~18, no.~2, pp. 1413--1452, 2016.

\bibitem{Liu2016Charging}
Q.~Liu, J.~Wu, P.~Xia, S.~Zhao, W.~Chen, Y.~Yang, and L.~Hanzo, ``Charging
  unplugged: Will distributed laser charging for mobile wireless power transfer
  work?'' \emph{IEEE Veh. Technol. Mag.}, vol.~11, no.~4, pp. 36--45, Dec.
  2016.

\bibitem{2018Charging}
V.~Iyer, E.~Bayati, R.~Nandakumar, A.~Majumdar, and S.~Gollakota, ``Charging a
  smartphone across a room using lasers,'' \emph{Proceedings of the Acm on
  Interactive Mobile Wearable \& Ubiquitous Technologies}, vol.~1, no.~4, pp.
  1--21, 2018.

\bibitem{2019Two}
X.~Zhang, J.~Grajal, J.~Luis Vazquez-Roy, U.~Radhakrishna, X.~Wang, W.~Chern,
  L.~Zhou, Y.~Lin, P.~C. Shen, and X.~Ji, ``Two-dimensional mos2-enabled
  flexible rectenna for wi-fi-band wireless energy harvesting,'' \emph{Nature},
  vol. 566, no. 7744, pp. 368--372, 2019.

\bibitem{Wangw}
W.~Wang, Q.~Zhang, H.~Lin, M.~Liu, X.~Liang, and Q.~Liu, ``Wireless energy
  transmission channel modeling in resonant beam charging for iot devices,''
  \emph{IEEE Internet of Things Journal}, vol.~6, no.~2, pp. 3976--3986, Jan.
  2019.

\bibitem{mobility}
M.~Liu, H.~Deng, Q.~Liu, J.~Zhou, M.~Xiong, L.~Yang, and G.~B. Giannakis,
  ``Simultaneous mobile information and power transfer by resonant beam,''
  \emph{IEEE Transactions on Signal Processing}, vol.~69, pp. 2766--2778, May
  2021.

\bibitem{safety}
W.~Fang, H.~Deng, Q.~Liu, M.~Liu, Q.~Jiang, L.~Yang, and G.~B. Giannakis,
  ``Safety analysis of long-range and high-power wireless power transfer using
  resonant beam,'' \emph{IEEE Transactions on Signal Processing}, vol.~69, pp.
  2833--2843, May 2021.

\bibitem{testbed}
Q.~Liu, M.~Xiong, M.~Liu, Q.~Jiang, W.~Fang, and Y.~Bai, ``Mobile wireless
  power transfer using a self-aligned resonant beam,'' 2021.

\bibitem{2fang2018}
W.~Fang, Q.~Zhang, M.~Liu, Q.~Liu, and P.~Xia, ``Earning maximization with
  quality of charging service guarantee for {I}o{T} devices,'' \emph{IEEE
  Internet Things J.}, vol.~6, no.~1, pp. 1114--1124, Feb. 2019.

\bibitem{Hangguo}
J.~Lim, T.~S. Khwaja, and J.~Ha, ``Wireless optical power transfer system by
  spatial wavelength division and distributed laser cavity resonance,''
  \emph{Opt. Express}, vol.~27, no.~12, pp. A924--A935, Jun 2019.

\bibitem{Sheng21}
Q.~Sheng, M.~Wang, H.~Ma, Y.~Qi, J.~Liu, D.~Xu, W.~Shi, and J.~Yao,
  ``Continuous-wave long-distributed-cavity laser using cat-eye
  retroreflectors,'' \emph{Opt. Express}, vol.~29, no.~21, pp.
  34\,269--34\,277, Oct 2021.

\bibitem{liu2022large}
J.~Liu, A.~Wang, Q.~Sheng, Y.~Qi, S.~Wang, M.~Wang, D.~Xu, S.~Fu, W.~Shi, and
  J.~Yao, ``Large-range alignment-free distributed-cavity laser based on an
  improved multi-lens retroreflector,'' \emph{Chinese Optics Letters}, vol.~20,
  no.~3, p. 031407, 2022.

\bibitem{chargernetwork}
{Xiao Lu}, D.~{Niyato}, {Ping Wang}, {Dong In Kim}, and {Zhu Han}, ``Wireless
  charger networking for mobile devices: fundamentals, standards, and
  applications,'' \emph{IEEE Wireless Commun.}, vol.~22, no.~2, pp. 126--135,
  Apr. 2015.

\end{thebibliography}

\section*{Biographies}
\vspace{-13 mm}
\begin{IEEEbiographynophoto}{Mingqing Liu} 
\small
	\setlength{\baselineskip}{10pt}
	(clare@tongji.edu.cn) received the B.S. degree in computer science and technology from the Northwest A\&F University, Yangling, China, in 2018. She is currently pursuing the Ph.D. degree with the College of Electronics and Information Engineering, Tongji University, Shanghai, China. Her research interests lie in the areas of wireless power transfer, development of remote wireless charging technology, and the Internet of Things.

\end{IEEEbiographynophoto}

\vspace{-2 mm}
\begin{IEEEbiographynophoto}{Qingwen Liu}
	\small
	\setlength{\baselineskip}{10pt}
(M'07--SM'15) received the B.S. degree in electrical engineering and information science from the University of Science and Technology of China, Hefei, China, in 2001 and the M.S. and Ph.D. degrees from the Department of Electrical and Computer Engineering, University of Minnesota, Minneapolis, MN, USA, in 2003 and 2006, respectively. He is currently a professor with the College of Electronics and Information Engineering, Tongji University, Shanghai, China.
His research interests lie in the areas of wireless power transfer and Internet of Things. He is a Senior Member of the IEEE.
\end{IEEEbiographynophoto}
\vspace{-2 mm}

\begin{IEEEbiographynophoto}{Hao Deng}
	\small
	\setlength{\baselineskip}{10pt} (denghao1984@tongji.edu.cn) received his B.S. and Ph.D. degrees from the Department of Physical Electronics, University of Electronic Science \& Technology, Chengdu, China, in 2007 and 2015, respectively. He is currently an Assistant Professor with the School of Software Engineering, Tongji University, Shanghai, China. His research interests focus on the areas of optical critical dimension measurement for semiconductors, wireless power transfer and Internet of Things.

\end{IEEEbiographynophoto}
\vspace{-2 mm}

\begin{IEEEbiographynophoto}{Mingliang Xiong}
	\small
	\setlength{\baselineskip}{10pt}
 (xiongml@tongji.edu.cn) received the B.S. degree in communication engineering from the Nanjing University of Posts and Telecommunications, Nanjing, China, in 2017. He is currently pursuing the Ph.D. degree with the College of Electronics and Information Engineering, Tongji University, Shanghai, China. His research interests include optical wireless communications, wireless power transfer, and the Internet of Things.
\end{IEEEbiographynophoto}

\vspace{-2 mm}

\begin{IEEEbiographynophoto}{Xinhe Wang}
	\small
	\setlength{\baselineskip}{10pt}
 (xinhelz1007@gmail.com) received the B.S. degree in Computer Science and Technology from Xi'an Jiaotong University, Xi'an, China, in 2017, and the M.S. Degree from Donald Bren School of Information \& Computer Science, University of California, Irvine, CA, USA, in 2019. She is currently pursuing the Ph.D. degree with the College of Electronics and Information Engineering, Tongji University, Shanghai, China. 
\end{IEEEbiographynophoto}

\end{document}